 \documentclass[aps,prb,twocolumn,superscriptaddress]{revtex4}

\usepackage[dvips]{graphicx}
\usepackage{times,bm,amsmath,amssymb}

\newcommand{\bQ}{\mbox{\boldmath $Q$}}

\newcommand{\bR}{\mbox{\boldmath $R$}}
\newcommand{\bk}{\mbox{\boldmath $k$}}

\begin{document}

\title{Optical Absorption Study by {\em Ab initio} Downfolding Approach:  
       Application to GaAs}

\author{Kazuma Nakamura} 
\thanks{Electronic mail: kazuma@solis.t.u-tokyo.ac.jp}
\affiliation{Department of Applied Physics, 
 University of Tokyo, 7-3-1 Hongo, Bunkyo-ku, Tokyo 113-8656, Japan}
 
\author{Yoshihide Yoshimoto} 
\affiliation{Institute for Solid State Physics, 
 University of Tokyo, 5-1-5 Kashiwanoha, Kashiwa, Chiba 277-8581, Japan} 

\author{Ryotaro Arita} 
\affiliation{Condensed Matter Theory Laboratory,
 RIKEN, 2-1 Hirosawa, Wako, Saitama 351-0198, Japan} 

\author{Shinji Tsuneyuki}
\affiliation{Department of Physics, 
 University of Tokyo, 7-3-1 Hongo, Bunkyo-ku, Tokyo 113-0033, Japan} 

\author{Masatoshi Imada}
\affiliation{Department of Applied Physics,
 University of Tokyo, 7-3-1 Hongo, Bunkyo-ku, Tokyo 113-8656, Japan}
\affiliation{JST, CREST,
 7-3-1 Hongo, Bunkyo-ku, Tokyo 113-8656, Japan}

\date{\today}

\begin{abstract}
 We examine whether essence and quantitative aspects of electronic excitation spectra 
 are correctly captured by an effective low-energy model constructed from an {\em ab initio} downfolding 
 scheme. 
 A global electronic structure is first calculated by {\em ab initio} density-functional calculations
 with the generalized gradient approximation. 
 With the help of constrained
 density functional theory,
 the low-energy effective Hamiltonian
 for bands
 near the Fermi level is 
 constructed
 by the downfolding procedure 
 in the basis of maximally localized Wannier functions.
 The excited states of this low-energy effective Hamiltonian ascribed to an extended Hubbard model 
 are calculated by using a low-energy solver. 
 As the solver, we employ 
 the Hartree-Fock approximation supplemented by 
 the single-excitation configuration-interaction method 
 considering electron-hole interactions.  
 The present three-stage method is applied to GaAs, 
 where eight bands are retained in the effective model after 
 the downfolding. 
 The resulting spectra well 
 reproduce the experimental results, indicating 
 that our downfolding scheme offers a satisfactory framework of 
 the electronic structure calculation, particularly
 for the excitations and dynamics as
 well as for the ground state.

\end{abstract}

\pacs{71.35Cc, 78.20.Bh, 78.40.Fy}

\maketitle
 
\section{Introduction}
 First-principles electronic-structure calculations based on 
 density-functional theory~\cite{Ref_DFT}(DFT) within the local density 
 approximation (LDA) or the generalized gradient approximation (GGA) 
 for the exchange-correlation (XC) functional 
 have opened a way to predict ground-state properties 
 of various materials without introducing {\it ad hoc} parameters.
 However, there exist serious problems 
 in which the DFT fails even qualitatively. 
 Typical examples are found in strongly-correlated electron systems  
 such as the genuine Mott insulator, where an insulating gap opens 
 in partially filled bands solely
 owing to the strong local electron-electron repulsion.\cite{Ref_Mott}          
 The DFT with LDA/GGA often predicts metals
 for these systems,\cite{Ref_Metal}            
 indicating the fact that the XC functionals based on 
 LDA/GGA do not correctly capture the local correlation 
 in real space. 
 Other typical example is found in dynamics and excitation spectra of 
 electrons, in which many-body correlation effects are also 
 essential.\cite{Ref_Hanke,Ref_Toyozawa}  
 Even semiconductors, being supposed to belong to weakly-correlated 
 electron systems in the ground state, may have highly-degenerate           
 excited states arising from the local electron correlation effects, 
 and thus the single-particle approximations 
 such as the Kohn-Sham~\cite{Ref_KS} and
 Hartree-Fock~\cite{Ref_HF} schemes 
 break down in general.  
 It is well known that incorporating two-particle interactions  
 between electrons and holes generated by the excitation 
 is crucial in describing the 
 electronic structure at low-energy 
 levels.
 A typical example is found in excitonic
 excitations.\cite{Ref_Hanke,Ref_Toyozawa,Ref_BSE,Ref_TDDFT,Ref_Louie,Ref_OA_THEORY}   

 To treat these excitations properly, we clearly need to go beyond 
 the single-particle theory, while 
 a full {\em ab initio} calculation taking into account 
 the many-body correlation effects is practically intractable. 
 To go beyond the LDA/Hartree-Fock levels, we are required to 
 develop a sufficiently accurate but efficient and practically 
 feasible method.  This challenge, so-called ``beyond LDA/Hartree-Fock'' 
 problem has attracted 
 growing interest.\cite{Ref_GW,Ref_SCISSORS,LDA+U,Ref_LDA+DMFT,
 Aryasetiawan2004,Solovyev2,Solovyev,Ref_LDA+PIRG2,Ref_DQMC,Ref_TC}
 The GW method~\cite{Ref_GW,Ref_SCISSORS} has been developed to incorporate self-energy effects basically
 on the level of the random phase approximation (RPA) while
 strong correlation and fluctuation effects beyond the RPA level require a more accurate and reliable
 treatment.   
 Especially, an {\em ab initio} three-stage scheme has been rapidly developed by
 combining two procedures, namely, LDA/Hartree-Fock framework and accurate low-energy 
 solvers.\cite{Ref_LDA+DMFT,Aryasetiawan2004,Solovyev2,Solovyev,Ref_LDA+PIRG2}
 The global electronic structure is first obtained by the LDA/Hartree-Fock scheme.
 In the next stage, one performs a bridging treatment,  
 that is {\em downfolding},~\cite{Ref_LDA+DMFT,Ref_LDA+PIRG2,Solovyev2,Solovyev}
 by eliminating the high-energy degrees of freedom
 leaving the low-energy effective model (Hamiltonian or Lagrangian) for 
 local bases like Wannier functions.\cite{Ref_MLWF,Ref_WF}   
 The downfolding determines parameters for the effective low-energy model 
 via first-principles calculations.  
 The resulting low-energy model is, in the final stage, solved by 
 low-energy reliable solvers such as dynamical mean field theory,\cite{Ref_LDA+DMFT} 
 path-integral renormalization group,~\cite{Ref_LDA+PIRG,Ref_LDA+PIRG2}       
 and/or various Monte-Carlo methods~\cite{Ref_MC}
 developed for treating the correlation effects. 
 Such a hierarchical three-stage scheme instead of a full {\em ab initio} calculation 
 allows us to perform a first-principles and parameter-free prediction
 of the electronic structure of the strongly-correlated electron system 
 within the present feasibility of computer.  
 
 In this paper, we present theoretical studies on the {\em ab initio} 
 downfolding scheme to assess the reliability for treating dynamical properties. 
 In our scheme, maximally localized Wannier functions are
 introduced as a basis function for representing the model Hamiltonian. 
 This basis offers computationally convenient choice, 
 because this Wannier function can be computed with any basis functions 
 (plane wave,\cite{Ref_MLWF}
 linearized muffin-tin orbital,\cite{Ref_MLWF-LMTO}
 linearized augmented plane wave,\cite{Ref_MLWF-FLAPW} etc). 
 Transfer parameters are evaluated by calculating Kohn-Sham 
 matrix elements in this basis, and 
 onsite/offsite 
 interaction parameters including screening effects  
 are determined via constrained
 calculations.\cite{Ref_CLDA,Solovyev,Solovyev2,Ref_GW_FOR_U,Aryasetiawan2004}  

 In the three-stage scheme, the reliability of the downfolding 
 procedure and the accuracy of the resulting model parameters
 are crucially important. In particular, the reliability in describing dynamics and
 excitation spectra has to be critically tested.
 For this purpose, 
 we make a critical comparison between experimental results and 
 computational results for the generated model. 
 In the present study, we focus on optical-absorption properties. 
 It is widely recognized and accepted in the literature that 
 the optical absorption of solids,
 in particular for semiconductors or insulators, is deeply affected
 by excitonic effects.\cite{Ref_Hanke,Ref_Toyozawa}            
 This effect originates from an effective Coulomb interaction
 between electrons and holes        
 and therefore is sensitive to 
 the magnitude of the interaction parameters 
 in the model Hamiltonian. 
 To examine this effect through the present formalism, we choose GaAs  
 as a representative material exhibiting spectral 
 enhancement due to the excitonic effect and 
 calculate its optical spectra by taking account of the 
 electron-hole interaction. 
 There exist many experimental~\cite{Ref_OA_EXPT_1,Ref_OA_EXPT_2}          
 and highly-accurate {\em ab initio}~\cite{Ref_Louie,Ref_OA_THEORY}           
 spectral data for this material. 
 Therefore, our downfolding formalism and determined parameters 
 can be critically tested by examining whether our model spectrum 
 reproduces those data satisfactorily. 

 This paper is organized as follows: 
 In Sec.~II we describe our downfolding procedure; 
 we introduce a complete-neglect-differential-overlap 
 model which is used as our target model Hamiltonian and 
 describe computational details for determining the
 model parameters. 
 In Sec.~III, to take into account the electron-hole interaction, 
 we introduce a single-excitation-configuration-interaction framework 
 for calculating an optical absorption. 
 Efficient techniques to evaluate one-body velocity matrix elements  
 needed in the spectral calculation, based on the 
 Wannier interpolation scheme, is described in appendix.  
 The calculated optical spectra 
 are compared with the experimental results. 
 Concluding remarks are given in Sec.~IV. 

\section{Downfolding Procedure}

\subsection{Global electronic structure by DFT}
\begin{figure}[h]
\includegraphics[width=0.40\textwidth]{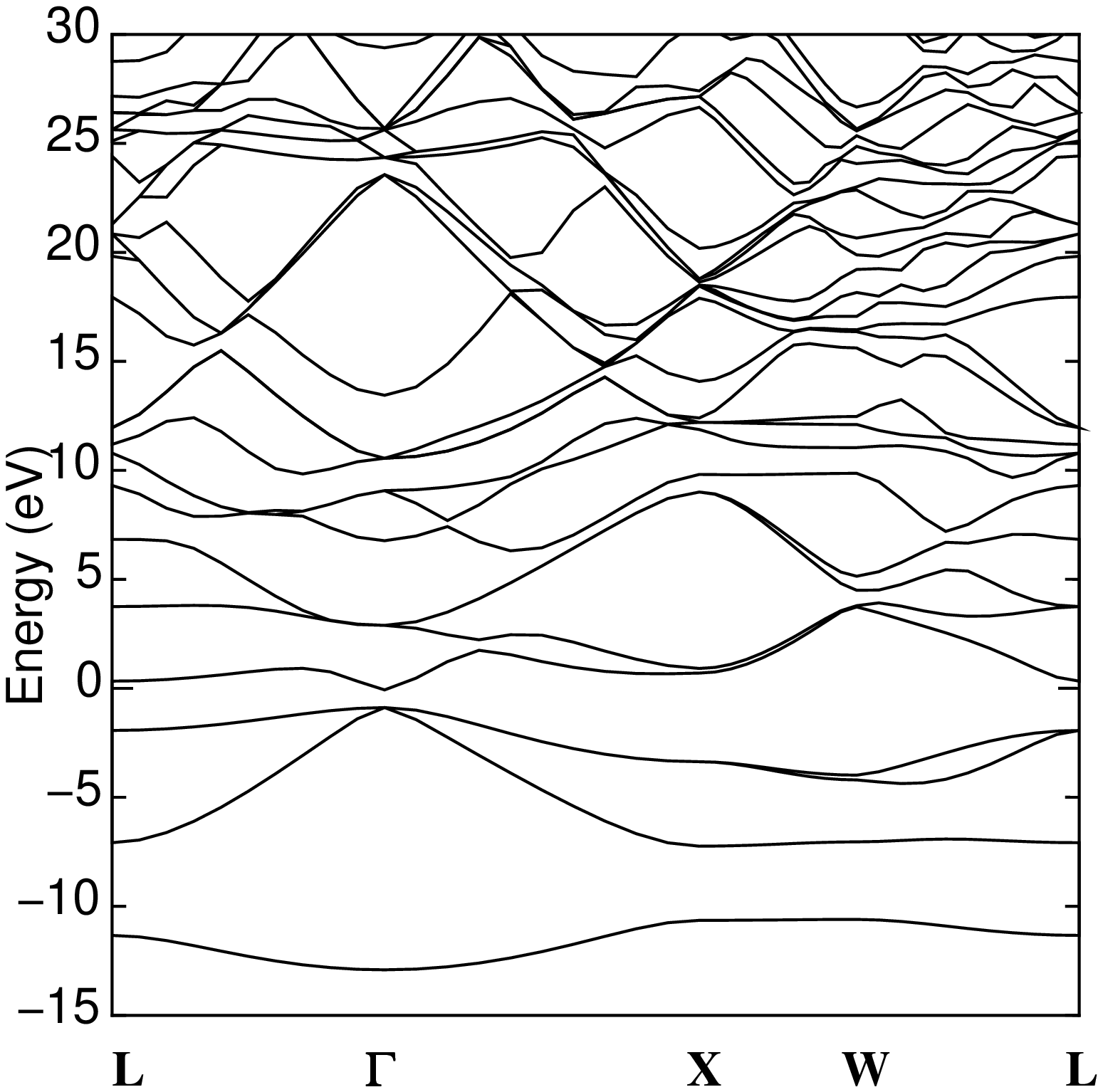} \ \ \ \ \ \ \ \ \ \ \ \ \  \\ 
\includegraphics[width=0.44\textwidth]{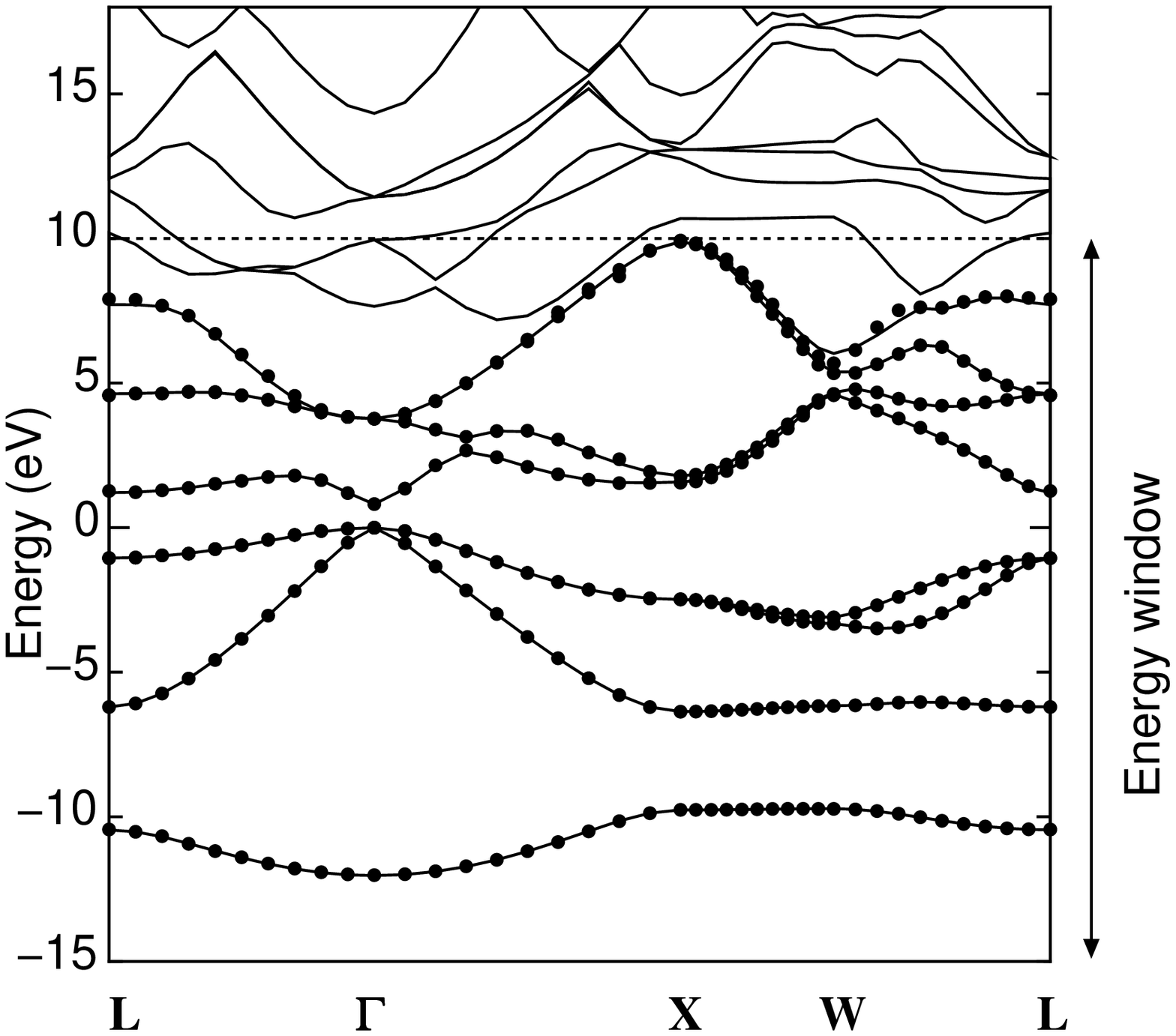}
\caption{(Top) 
 A global {\em ab initio} band structure of GaAs at        
 [$-15$ eV:$+30$ eV]. 
 (Bottom) {\em ab initio} original (solid line) and interpolated (dots) bands.
 Energy zero is set to the top of the valence bands. 
 The energy window is 
 set to [$-15$ eV:$+10$ eV].} 
\label{fig:band:global} 
\end{figure}

 The first procedure derives the global electronic band structure 
 by a conventional DFT scheme.
 The present scheme is based on {\em ab initio} density functional calculations with 
 {\em Tokyo Ab initio Program Package}~\cite{Ref_TAPP} developed 
 by the condensed-matter-theory group
 in the University of Tokyo.  
 With this code, band calculations 
 have been performed within the generalized gradient 
 approximation~\cite{Ref_PBE96} to the exchange correlation functional, 
 using a plane-wave basis set
 and the Troullier-Martins
 norm-conserving pseudopotentials~\cite{Ref_PP1} 
 in the Kleinman-Bylander representation.\cite{Ref_PP2} 
 The energy cutoff is set to 25 Ry, 
 and a 15$\times$15$\times$15 $k$-point sampling 
 is employed to represent electronic
 structures of the system. 
 The resulting global band structure of GaAs at an energy region [$-$15 eV:+ 30 eV] 
 is illustrated in the top panel of Fig.~\ref{fig:band:global} 

\subsection{Complete neglect differential overlap model} 
 Now we go onto the second stage and start the derivation of an effective low-energy Hamiltonian by
 the downfolding procedure. 
 Before going to the downfolding itself, we first specify the form of the final effective low-energy
 Hamiltonian. 
 In the present downfolding procedure, we will make several approximations
 by simplifying the low-energy effective Hamiltonian, in which we expect
 that the approximations do not alter the optical spectra seriously. 
 The first approximation is to consider only the diagonal Coulomb interaction
 and ignore the offdiagonal part (exchange interaction) in the low-energy effective model,
 which results in an extended Hubbard model or, in other words, the
 complete-neglect-differential-overlap (CNDO) model. 
 The CNDO Hamiltonian has originally been introduced by Pople~\cite{Ref_CNDO_1}  
 to study electronic structures of small organic molecules   
 and, to date, has been extended  
 to study various properties of 
 complicated systems ranging from
 transition-metal compounds~\cite{Ref_CNDO_2}
 to proteins~\cite{Ref_CNDO_3}
 and DNA.\cite{Ref_CNDO_4}   
 A remarkable property of this Hamiltonian is that 
 it considers all the degrees of the freedom of 
 valence electrons of the system,  
 which allows describing the individual 
 characters of the real material. 
 
 The crystal CNDO Hamiltonian $\mathcal{H}$ 
 consists of a one-body part $\mathcal{H}_{t}$ 
 and an interaction part $\mathcal{H}_{C}$:   
\begin{eqnarray}
   \mathcal{H}  
 = \mathcal{H}_{t}  
 + \mathcal{H}_{C}.   
  \label{eq:CNDO}  
\end{eqnarray}
The one-body part $\mathcal{H}_{t}$ is given by    
\begin{eqnarray}
  \mathcal{H}_{t} &=& \sum_{\sigma} \biggl\{ 
  \sum_{{\bm R}} \sum_{i} \sum_{\mu}
  I_{\mu i} a_{\mu i {\bm R}}^{\sigma \dagger}
  a_{\mu i {\bm R}}^{\sigma} \nonumber \\  
  &+& \sum_{{\bm R} {\bm R'}} \sum_{i j} \sum_{\mu \nu}  
  t_{\mu i \nu j}(\bR'-\bR) a_{\mu i {\bm R}}^{\sigma \dagger}
  a_{\nu j {\bm R'}}^{\sigma} \biggr\},               
  \label{eq:H_t}  
\end{eqnarray}
 where $a_{\mu i {\bm R}}^{\sigma \dagger}$ 
 ($a_{\mu i {\bm R}}^{\sigma}$) is a creation (annihilation) operator  
 of a valence electron with spin $\sigma$ in $\mu$-type 
 localized basis centered at $i$th site in lattice $\bR$. 
 As mentioned in the introduction, 
 we use the maximally localized Wannier function (MLWF) as 
 the basis function for representing the CNDO Hamiltonian. 
 In the present study of GaAs, 
 there are eight WF's 
 in the primitive cell,   
 where the first four 
 belong to a Ga site, and the remaining four 
 belong to an As site.
 Thus, the index $\mu$ specifies four types of
 lobe directions (band indices) of the MLWF's, 
 and the suffix $i$ specifies the Ga or As sites. 
 $I_{\mu i}$ 
 and $t_{\mu i \nu j} (\bR-\bR')$ 
 are the ionization potential 
 and the transfer integral, respectively.
 Notice that the translational symmetry in the crystal
 is explicitly considered for matrix elements; 
 $I_{\mu i {\bm R}} =$  
 $I_{\mu i}$ 
 and 
 $t_{\mu i {\bm R} \nu j {\bm R}'}$ 
 $=t_{\mu i \nu j}({\bm R}'-{\bm R})$            
 for any ${\bm R}$ and ${\bm R}'$.                

The interaction part $\mathcal{H}_{C}$ is written as 
\begin{eqnarray}
  \mathcal{H}_{C} &=& 
  \sum_{{\bm R}} \biggl\{ \sum_{i} \sum_{\mu}  
  U_{i} N_{\mu i {\bm R}}^{\uparrow} N_{\mu i {\bm R}}^{\downarrow}  
  + \sum_{i} \sum_{\mu < \nu}
  U'_{i} N_{\mu i {\bm R}} N_{\nu i {\bm R}} \biggr\} \nonumber \\ 
  &+& \sum_{{\bm R} {\bm R'}} \sum_{i j}
  V_{ij}(\bR-\bR') (N_{i {\bm R}} - Z_{i}) (N_{j {\bm R'}}-Z_{j}).  
  \label{eq:H_C}  
\end{eqnarray}
 Here, $N_{\mu i {\bm R}}^{\sigma}=a_{\mu i {\bm R}}^{\sigma \dagger} a_{\mu i {\bm R}}^{\sigma}$,  
 $N_{\mu i {\bm R}}=\sum_{\sigma} N_{\mu i {\bm R}}^{\sigma}$, and 
 $N_{i {\bm R}}=\sum_{\mu} N_{\mu i {\bm R}}$ are the number operators, and 
 $Z_{i}$ is the core charge. 
 $U_{i}$ and $U'_{i}$ are  
 the onsite intraorbital 
 and interorbital Coulomb repulsions, respectively. 
 $V_{ij} (\bR-\bR')$ in the third term is an interatomic Coulomb repulsion, and 
 it is assumed to be independent of the lobe directions $\mu$ and $\nu$. 

\subsection{Parameterization}
 We now describe the downfolding procedure and 
 parameterization for the CNDO model of Eq.~(\ref{eq:CNDO}).
 The downfolding consists of two parts.
 The first is the derivation of the kinetic energy part,
 where a tight-binding Hamiltonian ${\cal H}_t$ is derived. The second part is the 
 derivation of the interaction part ${\cal H}_C$.

\subsubsection{Kinetic Energy}
 The tight-binding Hamiltonian ${\cal H}_t$ given in Eq.~(\ref{eq:H_t}) is derived from 
 the global band structure after eliminating higher-energy bands. This downfolding may be performed by the 
 perturbation scheme.\cite{Solovyev,Ref_LDA+PIRG2,Solovyev2}
 The resultant band structure is normally very close to the low-energy part of 
 the original band structure and the difference is not discernible when the low-energy retained part is   
 isolated from the eliminated high-energy
 bands.\cite{Solovyev,Ref_LDA+PIRG2,Solovyev2} This means that the self-energy of the retained bands  
 caused by the higher-energy eliminated electrons is negligible. 
 Since such self-energy effects are smaller even in semiconductor systems, in this paper, 
 we employ the low-energy part of the bands as the retained bands after the elimination of the 
 higher-energy bands.

\begin{figure}[h]
\includegraphics[width=0.4\textwidth]{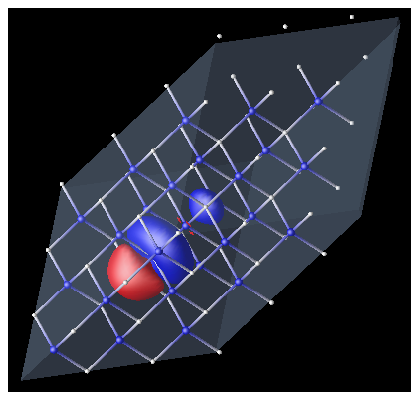}
\includegraphics[width=0.4\textwidth]{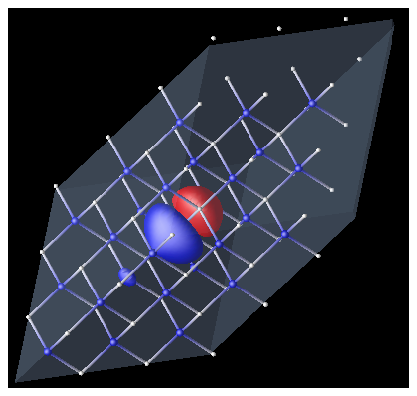}
\caption{(Color online) Maximally localized Wannier functions of Ga (top) and As (bottom).
 The amplitudes of the contour surface are
 $+ 0.5 / \sqrt{v}$ (blue) and 
 $- 0.5 / \sqrt{v}$ (red), where $v$ is the volume of the primitive cell.
 The shaded sheet represents a 3$\times$3$\times$3 fcc lattice
 and Ga and As nuclei are illustrated as
 gray and blue dots.}
\label{fig:MLWF} 
\end{figure}

 Now we retain eight bands near the Fermi level and construct Wannier orbitals from the 
 retained band structure.
 To this end, {\em ab initio} MLWF's 
 are constructed 
 with the Souza-Marzari-Vanderbilt algorithm.\cite{Ref_MLWF}  
 We set an energy window in the interval [$-15$ eV:$+10$ eV], 
 which includes four valence and four conduction bands of the system.  
 The resulting Ga and As MLWF's are displayed
 in the top and bottom panels of Fig.~\ref{fig:MLWF}, respectively. 
 We see that the Wannier functions are almost localized at a single site  
 and have an anisotropic character due to an $sp^{3}$ hybridization. 
 To show the accuracy of 
 low-energy band structures represented by the resultant WF's,  
 we compare in the bottom panel of Fig.~\ref{fig:band:global} original bands (solid line)
 with interpolated bands (dots) 
 obtained by diagonalizing $k$-space Kohn-Sham (KS) Hamiltonian
 matrices represented by the WF's. 
 We see a good agreement between the original
 and interpolated bands
 in the energy window. 
 Ionization potential $I_{\mu i}$ 
 and transfer integral $t_{\mu i \nu j}({\bm R})$ are extracted from the matrix 
 elements of the one-body KS Hamiltonian $\hat{h}_{\rm KS}$
 in the basis of the MLWF's $|w_{\mu i {\bm R}} 
 \rangle $as 
\begin{widetext} 
\begin{eqnarray} 
  I_{\mu i} &=& 
  \bigl\langle 
   w_{\mu i {\bm 0}} 
  \bigl| 
   \hat{h}_{{\rm KS}} 
  \bigr| 
   w_{\mu i {\bm 0}} 
  \bigr\rangle \ \ \ \ \ {\rm and}\ \ \ \ \  
  t_{\mu i \nu j}({\bm R}) =
  \bigl\langle 
   w_{\mu i {\bm 0}} 
  \bigl| 
   \hat{h}_{{\rm KS}} 
  \bigr| 
   w_{\nu j {\bm R}} 
  \bigr\rangle,     
  \label{eq:transfer}  
\end{eqnarray}
\end{widetext} 
 respectively.        
 Here ${\bm R} = l {\bm a}_1 + m {\bm a}_2 + n {\bm a}_3$ with  
 $-7 \le (l, m, n) \le +7$, and
 \{${\bm a}_1, {\bm a}_2, {\bm a}_3$\}  
 are primitive lattice vectors.    

\subsubsection{Interaction Energy}
 We next 
 derive the interaction 
 parameters $U_{i}$, $U'_{i}$, and $V_{ij}({\bm R})$
 for the low-energy model. 
 In the original CNDO framework, 
 the Coulomb interaction between electrons 
 has long-ranged tail as scaled by $1/r$ in the limit $r \to \infty$
 with $r$ being a distance between two electrons.      
 We are, however, interested in the electronic interaction in condensed phase 
 for which the Coulomb interaction is effectively screened. 
 In fact, the dielectric constant of GaAs is rather high;
 $\epsilon_0=10.6$ experimentally,\cite{Ref_GaAs_eps}      
 indicating that the Coulomb interaction
 decays as $1/(\epsilon_0 r)$
 in the limit $r \to \infty$.
 We thus employ an approximation for this interaction; 
 we keep only onsite interactions $U_{i}$ and $U'_{i}$
 and the nearest-neighbor interaction $V$. 
 
 These parameters are determined with a constrained DFT framework
 following a ``hopping-cutoff" treatment.~\cite{Ref_Held} 
 As the basic strategy, one first kinetically decouple 
 a specific site from the rest of the system, 
 thus leaving this site isolated as the so-called atomic limit.
 This decoupling treatment is achieved by switching off the hoppings 
 between the Wannier orbitals at the specific site and all the other 
 orbitals, where we identify the hoppings  
 with the off-diagonal Kohn-Sham matrix elements in the representation of the 
 Wannier functions. 
 With such a hopping cutoff, the standard constrained total energy
 calculations
 are performed to 
 generate a potential energy surface with respect to 
 constrained parameters such as occupancies of the 
 Wannier orbitals belonging to the decoupled site.  
 The interaction parameters obtained with
 quadratic fitting to the resulting potential energy data 
 include screening effects ascribed to 
 the relaxation of the valence electron density around the decoupled site. 
 In the present case, the procedure for determining the interaction 
 parameters is somewhat complicated because of a large number 
 of the interaction parameters to be derived. 
 So, in the practical work, we divide the treatment 
 into two steps; the determination of $U$ and $U'$ 
 and the subsequent step for determining 
 an offsite parameter $V$. 

 The basic strategy for obtaining $U$ and $U'$ 
 is to generate potential-energy-data sets
 with respect to two types of the constrained parameters;  
 (I) the first potential-energy data are obtained from the constrained calculations  
 with respect to occupancy $q_{\mu I}$ 
 of a specific Wannier orbital $\mu$ at the site $I$,  
 and (II) another data are obtained from the constrained calculations  
 for a site occupancy $Q_I$ 
 defined as 
 the total amount of the orbital occupancies
 belonging to this site; $\sum_{\mu} q_{\mu I}$.  
 The curvature of  
 the first potential energy curve 
 plotted as a function of the orbital occupation $q_{\mu I}$ 
 gives an estimate of the onsite intraorbital 
 interaction for the orbital $\mu$ at the site $I$,
 while the curvature of the second data 
 represents the averaged value 
 over the onsite intraorbital/interorbital interactions. 
 From quadratic fitting to the mixed two data, 
 we can determine the $U$ and $U'$ parameters reasonably (see below). 
 Though the types of the constraints are  
 different in the two calculations, 
 the method itself is the same, so,  
 here, we describe only details for the
 constrained calculation with respect to the single orbital occupancy. 

 The practical calculation proceeds as follows: 
 We first consider a $3\times3\times3$ fcc supercell containing 
 $54$ atoms, 
 \cite{Ref_supercell}  
 and choose the central Ga site placed at the origin as the decoupled site.
 (Here, we describe only the Ga case, but a parallel treatment 
 can be applied to the $U$ and $U'$ determination for the As site.)  
 We next introduce a cutting operator $\hat{\Lambda}_{cut}$ to 
 switch off the hopping integrals connecting the four Wannier orbitals 
 of this site to the other Wannier orbitals, 
\begin{eqnarray}
   \hat{\Lambda}_{cut} 
&=& - \hat{P}_{I {\bm 0}} \hat{h}_{KS} \hat{P}_{W}
 - \hat{P}_{W} \hat{h}_{KS} \hat{P}_{I {\bm 0}}
 + \hat{P}_{I {\bm 0}} \hat{h}_{KS} \hat{P}_{I {\bm 0}} \nonumber \\ 
&+& \sum_{\mu} \left| w_{\mu I {\bm 0}} \rangle 
   I_{\mu I}   
   \langle w_{\mu I {\bm 0}} \right| 
 \label{eq:cut},   
\end{eqnarray}
 where $\hat{h}_{KS}$ is a one-body KS Hamiltonian, and 
 $\hat{P}_{W}$ 
 is a projector onto the total Wannier orbitals, 
\begin{eqnarray}
   \hat{P}_{W}
 = \sum_{{\bm X}} \sum_{i} \sum_{\mu} 
   \left| w_{\mu i {\bm X}} \rangle   
   \langle w_{\mu i {\bm X}} \right|.  
\end{eqnarray}
 Here, ${\bm X}$ is a lattice vector denoting a supercell, 
 the suffix $i$ specifies the sites in the supercell,
 and $\mu$ stands for the band index of the Wannier orbital. 
 $\hat{P}_{I {\bm 0}}$ 
 in Eq.~(\ref{eq:cut}) 
 is a projector onto the Wannier orbitals belonging to 
 the decoupled $I$ site    
 in the home cell ($\bm{X}=\bm{0}$), 
\begin{eqnarray}
   \hat{P}_{I {\bm 0}}
 = \sum_{\mu} \left| w_{\mu I {\bm 0}} \rangle   
   \langle w_{\mu I {\bm 0}} \right|.  
\end{eqnarray}
 With the cutting operator 
 in Eq.~(\ref{eq:cut}),   
 we define 
 the constrained total energy as 
\begin{widetext}        
\begin{eqnarray}
  E_{{\rm ctot}}
 = \min_{\rho} 
 \Biggl\{ 
 \mathcal{F}[\rho] + \frac{1}{N} \sum_{\bm k} \sum_{\alpha} f_{\alpha {\bm k}}
 \bigl\langle \phi_{\alpha {\bm k}} 
 | \hat{\Lambda}_{cut}  | 
 \phi_{\alpha {\bm k}} \bigr\rangle  
 + \lambda_{\mu I {\bm 0}} \left[
 \frac{1}{N} \sum_{\bm k} \sum_{\alpha} f_{\alpha {\bm k}}
 | \bigl\langle w_{\mu I {\bm 0}} | \phi_{\alpha {\bm k}}
 \bigr\rangle |^{2}  
 - q_{\mu I {\bm 0}} 
 \right]  
 \Biggr\}. 
 \label{eq:E_CTOT}
\end{eqnarray}
\end{widetext} 
 Here, $\mathcal{F}[\rho]$ is a usual density functional with a total charge density 
 $\rho({\bm r})=\frac{1}{N}\sum_{\alpha {\bm k}} 
 f_{\alpha {\bm k}} | \phi_{\alpha {\bm k}} ({\bm r}) |^{2}$ with $N$ being 
 the total number of $k$ points, 
 the first term in the bracket $\left[ \cdots \right]$
 is the definition itself for the  
 orbital occupancy
 of the disconnected Wannier orbital $| w_{\mu I {\bm 0}} \rangle$,  
 and $\lambda_{\mu I {\bm 0}}$ is the Lagrange
 multiplier associated with the constraint to fix 
 the orbital occupancy at 
 $q_{\mu I {\bm 0}}$.
 A functional derivative of $E_{{\rm ctot}}$ with 
 respect to the Bloch orbital $\phi_{\alpha {\bm k}}$ leads to the 
 following constrained KS equation,  
\begin{eqnarray}
 \Biggl[ \hat{h}_{KS} + \hat{\Lambda}_{cut} 
 + \lambda_{\mu I {\bm 0}} | w_{\mu I {\bm 0}} \rangle 
  \langle w_{\mu I {\bm 0}} |  
 \Biggr]
 | \phi_{\alpha {
 \bf k}} \rangle =
 \epsilon_{\alpha {\bm k}} | \phi_{\alpha {\bm k}} \rangle, 
 \label{eq:KSeq}   
\end{eqnarray} 
 where the third term in the left hand side 
 is an additional potential due to the 
 occupancy constraint.
 For numerical details for solving the equation, 
 readers are referred to Ref.~29. 
 By solving the equation, we generate constrained potential energy data
 (we refer to these data as DATA I), and plot them 
 as functions of the orbital occupancy 
 $q_{\mu I {\bm 0}}$ and
 the site occupancy  
 $Q_{I{\bm 0}}=\sum_{\mu} q_{\mu I {\bm 0}}$.  
 In parallel to this treatment,
 the constrained calculations for 
 the site occupancy 
 are performed,
 where there is a small modification
 in the constrained KS equation (\ref{eq:KSeq}); 
 the additional potential
 is changed to 
 $ \lambda_{I {\bm 0}} \sum_{\mu} | w_{\mu I {\bm 0}} \rangle 
   \langle w_{\mu I {\bm 0}} |$.   
 We again monitor the calculated constrained total 
 energies as functions of 
 $q_{\mu I {\bm 0}}$ and 
 $Q_{I{\bm 0}}$ (DATA II).        
 With the resulting DATA I and II, 
 we perform quadratic fitting  
 of the following function around 
 $q_{\mu I{\bm 0}} = \bar{q}_{\mu I{\bm 0}}$ and 
 $Q_{I{\bm 0}} = \bar{Q}_{I{\bm 0}}$: 
\begin{eqnarray}
  f \left( Q_{I{\bm 0}}, q_{\mu I {\bm 0}} \right) 
 &=& \frac{1}{2} 
 U_{\rm Ga}
 \left( Q_{I{\bm 0}}-\bar{Q}_{I{\bm 0}} \right)^2 \nonumber \\
 &+& 2 \left( U'_{\rm Ga} - U_{{\rm Ga}}  \right)
 \left( Q_{I{\bm 0}}-\bar{Q}_{I{\bm 0}} \right)
 \left( q_{\mu I {\bm 0}} - \bar{q}_{\mu I {\bm 0}} \right) \nonumber \\ 
 &+& 2 \left( U_{{\rm Ga}}  - U'_{\rm Ga} \right)
 \left( q_{\mu I {\bm 0}} - \bar{q}_{\mu I {\bm 0}} \right)^2, 
 \label{eq:fitting}   
\end{eqnarray}
 where 
 $\bar{q}_{\mu I{\bm 0}}$ and 
 $\bar{Q}_{I{\bm 0}}$ are equilibrium occupancies
 taken from the global band structure 
 with no additional potential 
 ($\lambda_{\mu I {\bm 0}}=0$).
 The form of the fitting function in Eq.~(\ref{eq:fitting}) is derived by 
 exploiting the character of the fourfold degeneracy of the Wannier orbitals 
 $\left\{ w_{\mu I {\bm 0}} \right\}$ (Appendix A).
 The $U_{{\rm Ga}}$ and $U'_{\rm Ga}$ values thus obtained 
 are 2.39 eV and 2.17 eV, respectively,  
 which are largely reduced from the bare interaction 
 values $U_{\rm Ga}^{0}=9.25$ eV and $U'^{0}_{\rm Ga}=7.89$ eV. 
 The same procedure is applied to the 
 $U$ and $U'$ determinations of the As site. 
 It was found to be 
 $U_{\rm As} = 2.71$ eV and $U'_{\rm As} = 2.09$ eV. 
 The corresponding bare values 
 $U_{\rm As}^{0}$ and $U'^{0}_{\rm As}$ 
 are 11.45 eV and 9.80 eV, respectively.         

 We next describe the determination of the offsite parameter $V$. 
 The interaction depends on the relative configuration between 
 the Wannier orbitals.  
 For example, let us consider a configuration  
 formed by a pair of the Ga Wannier orbital 
 and the As Wannier orbital, 
 where these orbitals  
 face along the covalent bond of the two atoms
 (we call it {\em facing configuration}). 
 An example for the facing configuration 
 can be found in the two Wannier orbitals displayed in 
 the top and bottom panels of Fig.~\ref{fig:MLWF}.  
 Obviously, the strength of the Coulomb repulsion in the facing configuration
 is relatively large 
 compared to that in the other configurations. 
 It should be noted here that the $V$ parameter affects 
 {\em renormalized} transfer integrals   
 [see Eq.~(\ref{eq:Fock_R}) in Sec.~II~D 
 for the explicit form of the renormalized transfer integral]. 
 Since bonding and anti-bonding orbitals are formed by 
 the Wannier orbitals in the facing configuration, 
 the bonding and anti-bonding splitting, i.e., band gap itself,  
 is dominated by 
 the magnitude of a renormalized transfer integral between 
 the Wannier orbitals in the facing configuration.
 Thus, the $V$ value 
 in this configuration  
 is crucial  
 for an accurate description of the low-energy band structure
 and optical excitations.
 The contributions from the intersite interaction
 in the other configurations are much smaller in 
 optical response. 
 The result would not change 
 when we slightly overestimate them 
 by $V$ in the facing configuration, 
 because the renormalized part to the transfer integral 
 appears as the product of $V$ and a density matrix 
 [see Eq.~(\ref{eq:Fock_R})] and 
 it was found that the intersite density-matrix elements 
 are almost zero except for that of the facing configuration 
 in the present GaAs case. 
 Therefore, we calculate the $V$ value in the facing configuration 
 and employ it as the $V$ value of the CNDO model. 

 The actual determination of the $V$ parameter proceeds as follows: 
 We first choose two decoupled sites (the Ga site placed at the origin and 
 the neighboring As site being in the [111] direction).
 The similar cutting treatment to Eq.~(\ref{eq:cut})
 but extending the single-site-cutting formalism
 to the double-sites-cutting formalism 
 is applied for this purpose. 
 Then, we perform constrained calculations
 by imposing a constraint 
 that two occupancies of the Wannier orbitals 
 $| w_{\mu I {\bm 0}} \rangle$ and    
 $| w_{\nu J {\bm 0}} \rangle$ 
 in the facing configuration 
 are kept at  
 $q_{\mu I {\bm 0}}$ and 
 $q_{\nu J {\bm 0}}$, respectively.
 We then draw the two-dimensional potential energy surface with 
 respect to the constrained parameters 
 $q_{\mu I {\bm 0}}$ and 
 $q_{\nu J {\bm 0}}$, 
 and  perform a fitting of the quadratic function 
 $ \frac{1}{2} U_{{\rm Ga}} (q_{\mu I {\bm 0}} - \bar{q}_{\mu I {\bm 0}})^2 
 + \frac{1}{2} U_{{\rm As}} (q_{\nu J {\bm 0}} - \bar{q}_{\nu J {\bm 0}})^2 
 + V (q_{\mu I {\bm 0}} - \bar{q}_{\mu I {\bm 0}})
     (q_{\nu J {\bm 0}} - \bar{q}_{\nu J {\bm 0}})$
 to the potential energy data. 
 This fitting is performed by fixing $U_{\rm Ga}$ and $U_{\rm As}$ at 
 predetermined values in the preceding $U$ and $U'$ determination;
 we treat only $V$ as a single fitting parameter
 to avoid fitting errors and uncertainties. 
 The $V$ value thus determined is 0.71 eV, 
 where we again see the large reduction from the bare interaction value 
 $V^{0}=7.35$ eV. 
 The resulting interaction parameters are summarized in TABLE~\ref{PARAM}.
\begin{table}[h] 
\caption{Interaction parameters determined in the present 
 downfolding procedure.   
 The energy unit is eV.}

\ 

\begin{tabular}{cccccc} \hline \hline
  & $U_{\rm Ga}$  & $U'_{\rm Ga}$ & $U_{\rm As}$  & $U'_{\rm As}$ & $V$           \\ \hline 
  & 2.39\ \ \ \ \ & 2.17\ \ \ \ \    & 2.71\ \ \ \ \ & 2.09\ \ \ \ \    & 0.71\ \ \ \ \ \\ \hline \hline 
\end{tabular}
\label{PARAM} 
\end{table}

 In general, the downfolded model contains an energy dependence in the 
 interaction because the screening by the high-energy electrons necessarily 
 causes a non-Markoffian and retardation effect. 
 Such an energy dependence in the screened Coulomb interaction $W(\omega)$
 is not described by the effective Hamiltonian. 
 However, in the low-energy region, the screened Coulomb 
 interaction is normally saturated to a constant and represented by the limiting value at $\omega=0$.
 The constrained scheme is roughly regarded as the procedure to obtain this $\omega =0$ limit.
 The effect of the larger (less screened ) interaction $W$ at larger $\omega$
 as well as the effective interaction arising from 
 virtual transition to eliminated bands           
 can be accounted by the further consideration of the self-energy effect
 on the low-energy part.\cite{Aryasetiawan2004} 
 For a wide band system such as GaAs, however, this effect may not be large~\cite{Ref_Louie}
 and we thus ignore this effect.  

\subsection{Hartree-Fock approximation}

 In the electronic structure calculation of a typical semiconductor, GaAs, 
 we expect that the strong correlation effect does not appear 
 in the ground state 
 and the Hartree-Fock (HF) approximation provides us a reasonable result,
 although the excitation spectra are reliably determined only through the
 account of the correlation effect more accurately. 
 In this section, we consider how the HF solution for the ground state is calculated. 

 The CNDO HF Hamiltonian in the basis representation of MLWF's is written as 
\begin{eqnarray}
  \mathcal{H}_{{\rm HF}} &=& \sum_{\sigma} 
             \sum_{{\bm R} {\bm R'}} 
             \sum_{i j} 
             \sum_{\mu \nu} \ 
             F_{\mu i \nu j}(\bR'-\bR) 
             a_{\mu i {\bm R} }^{\sigma \dagger}
             a_{\nu j {\bm R'}}^{\sigma},  
  \label{eq:CNDO-HF}  
\end{eqnarray}
 where $F_{\mu i \nu j}(\bR'-\bR)$ is the Fock matrix or 
 the renormalized transfer matrix by the interaction part 
 $\mathcal{H}_{C}$ of Eq.~(\ref{eq:H_C}) and 
 the matrix element is given by             
\begin{widetext} 
\begin{eqnarray}
  F_{\mu i \nu j} (\bR'-\bR) 
  = \left\{
    \begin{array}{@{\,}ll}
      \mbox{%
        \parbox{7.5cm}{%
          $\displaystyle
          I_{\mu i} + \left[ \left( Q_{i}({\bm 0})-Z_{i} \right)
          - \frac{1}{2} \left( Q_{\mu i \mu i}({\bm 0}) - 2 \right) \right] U_{i} 
          $
          \\
          $\displaystyle
          + \sum_{ {\bm R}'' k} \left[ Q_{k}(\bR''-\bR) - Z_{k} \right] V_{ik}(\bR''-\bR)  
          $
        }
      } & \mbox{($\bR=\bR', i = j, \mu = \nu$),}
    \\[+10pt]
      \displaystyle 
      t_{\mu i \nu i}({\bm 0}) - \frac{1}{2} Q_{\mu i \nu i}({\bm 0}) U'_{i}   
      & \mbox{($\bR = \bR', i = j, \mu \ne \nu$),} 
    \\[+10pt]
      \displaystyle 
      t_{\mu i \nu j}({\bm 0}) - \frac{1}{2} Q_{\mu i \nu j}({\bm 0}) V_{ij}({\bm 0})  
      & \mbox{($\bR=\bR', i \ne j$),} 
    \\[+10pt]
      \displaystyle
      t_{\mu i \nu j}(\bR'-\bR)
      - \frac{1}{2} Q_{\mu i \nu j} (\bR'-\bR) 
      V_{ij} (\bR'-\bR)  
      & \mbox{($\bR \ne \bR'$),}
    \end{array}
  \right.
  \label{eq:Fock_R}  
\end{eqnarray}
\end{widetext} 
 Here $\bQ(\bR'-\bR)$ is the density matrix                
 and the matrix element is given by $
  Q_{\mu i \nu j} (\bR'-\bR) = 
  \bigl\langle \Phi_{{\rm HF}} \bigl| \sum_{\sigma}
  a_{\mu i {\bm R}}^{\sigma \dagger}
  a_{\nu j {\bm R'}}^{\sigma} 
  \bigr| \Phi_{{\rm HF}} \bigr\rangle, $
 where $|\Phi_{{\rm HF}}\rangle$ is the HF ground state.  
 The HF Hamiltonian can be diagonalized with the Bloch orbital, 
  $f_{\alpha \bk}^{\sigma \dagger} = 
  \left( 1/ \sqrt{N} \right)  
  \sum_{\mu i} C_{\mu i}^{\alpha}(\bk) 
  \sum_{{\bm R}} e^{i {\bm k}\cdot{\bm R}}
  a_{\mu i {\bm R}}^{\sigma \dagger}$, 
 where $\alpha$ and ${\bm k}$ are a band index and a wave vector, respectively,
 and $N$ is the total number of the unit cells in the system. 
 The coefficients $\{ C_{\mu i}^{\alpha}(\bk) \}$ are determined 
 by solving the following Hartree-Fock equation,  
\begin{eqnarray}
             \sum_{i j} 
             \sum_{\mu \nu} 
             F_{\mu i \nu j}(\bk) C_{\nu j}^{\alpha} (\bk) = 
             \epsilon_{\alpha \bk} C_{\mu i}^{\alpha} (\bk)              
  \label{eq:HFeq}  
\end{eqnarray}
 with 
\begin{eqnarray}
 F_{\mu i \nu j}(\bk) = 
 \frac{1}{N} \sum_{{\bm R}} F_{\mu i \nu j}(\bR)\ e^{i {\bm k}\cdot{\bm R}}. 
  \label{eq:Fock_k}  
\end{eqnarray}
 With the resulting $C_{\mu i}^{\alpha} ({\bm k})$,
 the real-space density matrix is calculated by 
\begin{eqnarray}
  Q_{\mu i \nu j} (\bR) = 
  \frac{1}{N} \sum_{{\bm k}}^{N} 
  Q_{\mu i \nu j} ({\bm k}) 
  e^{-i{\bm k}\cdot{\bm R}}  
  \label{eq:density_R}  
\end{eqnarray}
with 
\begin{eqnarray}
  Q_{\mu i \nu j} ({\bm k}) = 
  2 \sum_{\alpha}^{occ} 
  C_{\mu i}^{\alpha}({\bm k})
  C_{\nu j}^{\alpha *}({\bm k}).   
  \label{eq:density_k}  
\end{eqnarray}

 The CNDO total energy with the HF approximation is given by 
\begin{eqnarray}
  E_{{\rm tot}} &=& 
  \bigl\langle \Phi_{{\rm HF}} \bigl| 
  \mathcal{H}  
  \bigr| \Phi_{{\rm HF}} \bigr\rangle \nonumber \\
 &=& 
  \frac{1}{2} \sum_{{\bm k}} \sum_{i j} \sum_{\mu \nu} 
  Q_{\mu i \nu j} ({\bm k}) \nonumber \\ 
 &\times&  \biggl[ H_{\nu j \mu i}^{{\rm core}} ({\bm k}) 
        + F_{\nu j \mu i} ({\bm k}) \biggr],  
  \label{eq:CNDO_E}  
\end{eqnarray}
 where 
\begin{eqnarray}
  H_{\mu i \nu j}^{{\rm core}}(\bk) = 
   \frac{1}{N} \sum_{{\bm R}} H_{\mu i \nu j}^{{\rm core}} (\bR)
    e^{i {\bm k}\cdot{\bm R}}. 
  \label{eq:H_core}  
\end{eqnarray}
The matrix element of the core matrix           
$H_{\mu i \nu j}^{{\rm core}}(\bR)$ is written as 
\begin{widetext} 
\begin{eqnarray}
  H_{\mu i \nu j}^{{\rm core}} (\bR) 
  = \left\{
    \begin{array}{@{\,}ll}
      \mbox{%
        \parbox{8cm}{%
          $\displaystyle
          I_{\mu i} - (Z_{i}-1) U_{i} 
          - \sum_{ {\bm R}' j} Z_{j} V_{ij}(\bR')  
          $
        }
      } & \mbox{($\bR={\bm 0},i=j,\mu=\nu$),}
    \\[+5pt]
      \displaystyle t_{\mu i \nu j}(\bR)
      & \mbox{(otherwise).} 
    \end{array}
  \right.
\end{eqnarray}
\end{widetext} 

\begin{figure}[h]
\includegraphics[width=0.5\textwidth]{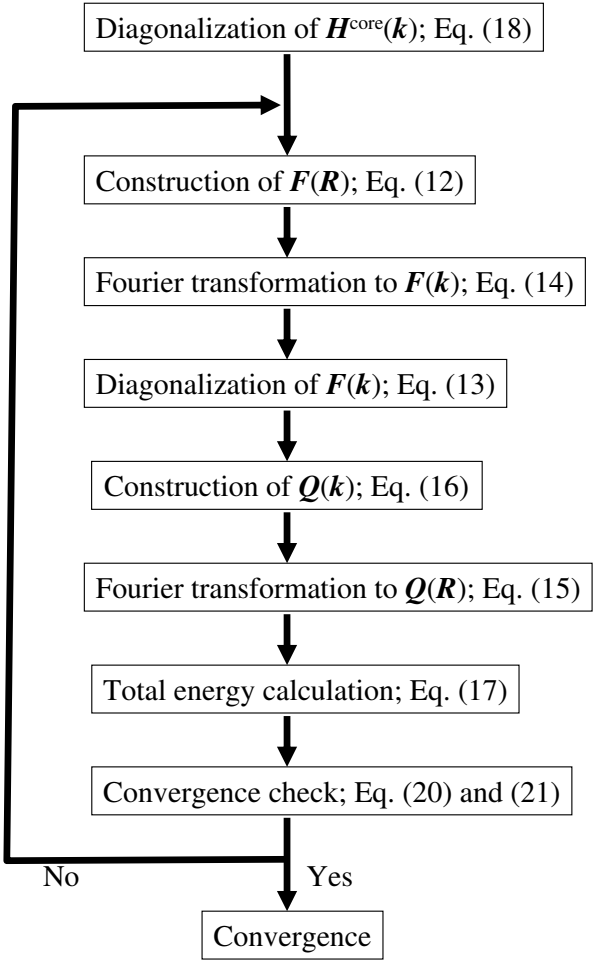} 
\caption{Schematic diagram for CNDO-HF band calculation}
\label{fig:flowchart} 
\end{figure}

 The actual CNDO HF calculation proceeds
 along the schematic diagram shown
 in Fig.~\ref{fig:flowchart}. 
 The initial density matrix 
 is made from diagonalizing the core matrix
 ${\bm H}^{{\rm core}}({\bm k})$ of Eq.~(\ref{eq:H_core}). 
 Since the Fock matrix ${\bm F}({\bm R})$ in Eq.~(\ref{eq:Fock_R})
 depends on the density matrix ${\bm Q}({\bm R})$ of Eq.~(\ref{eq:density_R}),
 the HF equation~(\ref{eq:HFeq}) 
 is solved self-consistently with an iterative procedure. 
 To check the convergence, we monitor a density-matrix difference, 
\begin{eqnarray}
  \delta Q^{(i)} = 
  \sqrt{ 
  \biggl( \frac{1}{8} \biggr)^{2} \sum_{i j} \sum_{\mu \nu}
  \biggl[ Q_{\mu i \nu j}^{(i)}({\bm 0})
   - Q_{\mu i \nu j}^{(i-1)}({\bm 0}) \biggr] },    
  \label{eq:delta_density}  
\end{eqnarray}
 and a total-energy difference,  
\begin{eqnarray}
  \delta E_{{\rm tot}}^{(i)}= 
  E_{{\rm tot}}^{(i)} - 
  E_{{\rm tot}}^{(i-1)},  
  \label{eq:delta_total}  
\end{eqnarray}
 where an upper suffix $i$ stands for the number of the iteration step. 
 The self-consistency condition we employ is satisfied when 
 $\delta Q^{(i)} \le 10^{-5}$ and   
 $\delta E_{{\rm tot}}^{(i)} \le 10^{-6}$ (a.u.).

\begin{figure}[h]
\includegraphics[width=0.4\textwidth]{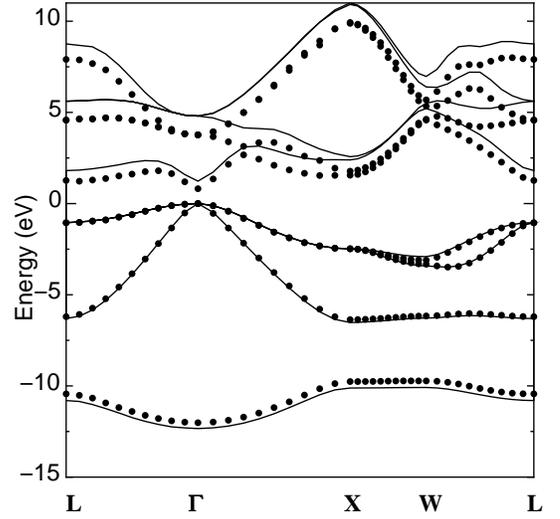}
\caption{
 Self-consistent GaAs CNDO-HF band structure (solid line)
 and {\em ab initio} interpolated band structure (dots).
 Energy zero is set to the top of the valence bands.}  
\label{fig:band:model} 
\end{figure}

 We show in Fig.~\ref{fig:band:model}    
 the self-consistent CNDO band structures (solid line), 
 together with 
 the {\em ab initio} interpolated band (dots).  
 We see a rigid band shift in the conduction band. 
 This rigid band shift is attributed to the renormalization 
 of the interaction part ${\mathcal H}_C$ in Eq.~(\ref{eq:H_C})
 into the one-body part; 
 the so-called self-energy correction considered within 
 the HF framework [see Eq.~(\ref{eq:Fock_R})]. 
 We note that the trend of the rigid band shift is basically the same as  
 the quasiparticle band shift observed in the GW calculation for
 semiconductor.\cite{Ref_GW,Ref_SCISSORS}

\section{Electronic excitation}
 In the previous section, we have described the procedure for the downfolding. 
 We now start calculating physical quantities using the downfolded model.
 The purpose of this sections is to examine the reliability of the 
 model obtained by the downfolding. 
 In particular, we highlight whether the model gives us a reliable excitation spectra
 and dynamical properties.
 For this purpose, we calculate optical-absorption spectra,
 based on a configuration-interaction (CI) treatment
 considering electron-hole interactions, and  
 compare the computational result with the experiments. 

\subsection{Single excitation configuration interaction}
 For electrons in solids, the number of configurations generated 
 by the electronic excitations is 
 in principle infinitely large, while the configurations 
 capable of practical computations are limited. 
 Since we are interested in optical processes, 
 we consider here only the single-excitation (SE) configurations  
 which play a primarily important role in 
 the linear-absorption process, because 
 the SE configurations directly couple with the HF ground state via 
 an electric dipole operator. 
 The calculation at the SECI level 
 in fact takes into account electron-hole interactions; 
 the so-called excitonic effect in the spectrum.  

The SECI many-body wave function with
 a wave vector ${\bm K}$ is written  
 as~\cite{Ref_Ikawa}     
\begin{eqnarray}
  \bigl| \Psi_{e {\bm K}} \bigr\rangle = 
  \sum_{{\bm k}}
  \sum_{a}^{occ} 
  \sum_{r}^{vir} 
  \tilde{C}_{a r {\bm k}}^{e {\bm K}}
  \bigl|^{1}\Psi_{a {\bm k}}^{r {\bm k}+{\bm K}} \bigr\rangle,   
  \label{eq:SECI}  
\end{eqnarray}
 where $|^{1}\Psi_{a {\bm k}}^{r {\bm k}+{\bm K}} \rangle$ is a spin-singlet 
 SE configuration given by 
\begin{eqnarray}
  \bigl|^{1}\Psi_{a {\bm k}}^{r {\bm k+K}} \bigr\rangle =    
  \frac{1}{\sqrt{2}} 
  \bigl( d_{r {\bm k}+{\bm K}}^{\uparrow \dagger} 
         d_{a {\bm k}}^{\uparrow} + 
         d_{r {\bm k}+{\bm K}}^{\downarrow \dagger} 
         d_{a {\bm k}}^{\downarrow} \bigr) 
  \bigl| \Phi_{{\rm HF}} \bigr\rangle.      
  \label{eq:SEC}  
\end{eqnarray}
 Here $d_{r {\bm k}}^{\sigma \dagger}$ 
     ($d_{a {\bm k}}^{\sigma}$) 
 is a creation (annihilation) operator of the Bloch electron 
 in a virtual $r$ (occupied $a$) band with spin $\sigma$ and 
 a wave vector ${\bm k}$. 
 The CI coefficients 
 \{$\tilde{C}_{a r {\bm k}}^{e {\bm K}}$\} 
 in Eq.~(\ref{eq:SECI}) and 
 the excitation energy $\Delta E_{e {\bm K}}$
 are obtained by solving the following CI equation,
\begin{eqnarray}
  \sum_{{\bm k'}} \sum_{b}^{occ} \sum_{s}^{vir} 
  A_{a r {\bm k}, b s {\bm k'}}^{\bm K} 
  \tilde{C}_{b s {\bm k'}}^{e \bm K} = 
  \Delta E_{e {\bm K}} 
  \tilde{C}_{a r {\bm k}}^{e \bm K}  
  \label{eq:SECIeq}  
\end{eqnarray}
with 
\begin{eqnarray}
  A_{a r {\bm k}, b s {\bm k'}}^{\bm K} &=&  
  \bigl\langle^{1} \Psi_{a {\bm k}}^{r {\bm k}+{\bm K}}  
  \bigl| {\cal{H}}- E_{{\rm HF}} \bigr|^{1}  
  \Psi_{b {\bm k'}}^{s {\bm k'}+{\bm K}} \bigr\rangle \nonumber \\ 
   &=& \bigl( \epsilon_{r {\bm k}+{\bm K}} - \epsilon_{a {\bm k}} \bigr) 
  \delta_{{\bm k}{\bm k'}} \delta_{a b} \delta_{r s} \nonumber \\ 
  &+& 2 \bigl\langle r_{{\bm k}+{\bm K}}  
                   a_{{\bm k}} 
      \big|  
                   b_{{\bm k'}}  
                   s_{{\bm k'}+{\bm K}} 
      \bigr\rangle  \nonumber \\  
  &-& \bigl\langle r_{{\bm k}+{\bm K}}  
                 s_{{\bm k'}+{\bm K}} 
    \big|  
                 b_{{\bm k'}}  
                 a_{{\bm k}} 
    \bigr\rangle,     
  \label{eq:AMAT}  
\end{eqnarray}
 where $E_{{\rm HF}}$ is the HF ground-state eigenenergy 
 for the many-body HF Hamiltonian $\mathcal{H}_{{\rm HF}}$ in Eq.~(\ref{eq:CNDO-HF});          
 $\mathcal{H}_{{\rm HF}} | \Phi_{{\rm HF}} \rangle 
 = E_{{\rm HF}} | \Phi_{{\rm HF}} \rangle$. 
 For the CNDO model, the second term in Eq.~(\ref{eq:AMAT}),
 called the exchange term, 
 is calculated as    
\begin{eqnarray}
 \bigl\langle r_{{\bm k}+{\bm K}}  
              a_{{\bm k}} 
 \big|  
              b_{{\bm k'}}  
              s_{{\bm k'}+{\bm K}} 
 \bigr\rangle &=&   
 \sum_{\mu i} 
 \sum_{\nu j} 
 C_{\mu i}^{r *} ({\bm k}+{\bm K}) 
 C_{\mu i}^{a} ({\bm k}) 
 C_{\nu j}^{b *} ({\bm k'}) \nonumber \\ 
 &\times& C_{\nu j}^{s} ({\bm k'}+{\bm K}) 
 {\mathcal V}_{\mu i \nu j}({\bm K}),  
  \label{eq:EX}  
\end{eqnarray}
 and the last term in Eq.~(\ref{eq:AMAT}),
 refereed to as the direct term, is evaluated by 
\begin{eqnarray}
 \bigl\langle r_{{\bm k}+{\bm K}}  
              s_{{\bm k'}+{\bm K}} 
 \big|  
              b_{{\bm k'}}  
              a_{{\bm k}} 
 \bigr\rangle &=&   
 \sum_{\mu i} 
 \sum_{\nu j} 
 C_{\mu i}^{r *} ({\bm k}+{\bm K}) 
 C_{\mu i}^{s} ({\bm k'}+{\bm K}) \nonumber \\ 
 &\times& C_{\nu j}^{b *} ({\bm k'}) 
 C_{\nu j}^{a} ({\bm k}) 
 {\mathcal V}_{\mu i \nu j}({\bm k}-{\bm k'}),  
  \label{eq:DR}  
\end{eqnarray}
 with
\begin{eqnarray}
 {\mathcal V}_{\mu i \nu j}(\bk) = 
 \frac{1}{N} \sum_{{\bm R}} 
 {\mathcal V}_{\mu i \nu j}(\bR)\ e^{i {\bm k}\cdot{\bm R}}   
\end{eqnarray}
 and  
\begin{eqnarray}
  {\mathcal V}_{\mu i \nu j} (\bR) 
  = \left\{
    \begin{array}{@{\,}ll}
      \displaystyle U_i 
      & \mbox{($\bR={\bm 0}, i=j, \mu=\nu$),}
    \\[+5pt]
      \displaystyle U'_{i}
      & \mbox{($\bR={\bm 0}, i=j, \mu \ne \nu$),}
    \\[+5pt]
      \displaystyle V_{ij}(\bR)
      & \mbox{(otherwise).} 
    \end{array}
  \right.
\end{eqnarray}
 Eqs.~(\ref{eq:EX}) and (\ref{eq:DR}) imply the repulsive-exchange 
 and attractive-Coulomb interactions between an electron in the $r$ and $s$ bands 
 and a positive hole in the $a$ and $b$ bands, with the total wave vector 
 being kept constant [$({\bm k}+{\bm K})-{\bm k}={\bm K}$].

 The structure of the present CNDO-HF-SECI equation~(\ref{eq:SECIeq}) 
 is basically the same as that of the Bethe-Salpeter equation 
 for two-particle Green's functions.\cite{Ref_BSE,Ref_Louie,Ref_OA_THEORY}            
 A difference is that the former electron-electron interaction
 in the two-electron integrals of Eqs.~(\ref{eq:EX}) and (\ref{eq:DR})                        
 is represented by ${\mathcal V}_{\mu i \nu j}({\bm R})$ 
 determined via the constrained scheme (see Sec.~II~C), while  
 the latter interaction is represented by 
 the screened Coulomb interaction evaluated with the 
 random phase approximation. 
 Computationally, we note that four-center Coulomb 
 integrals in the original exchange/direct terms  
 are reduced to two-center Coulomb integrals including only the sites $i$ and $j$  
 because of the CNDO approximation.\cite{Ref_CNDO_1}          
 Therefore the computational cost for the
 matrix evaluation is much smaller  
 in our CI calculation. 
 The most time consuming step
 is the 
 diagonalization of the CI matrix ${\bm A}^{{\bm K}}$, which is 
 scaled as the third power of the dimension of the 
 ${\bm A}^{{\bm K}}$;
 $( N \cdot N_{occ} \cdot N_{vir})^3$.

\subsection{Optical absorption}
 We next describe the expression for the SECI optical absorption, 
 which is given as the imaginary part of the macroscopic 
 transverse dielectric function,\cite{Ref_Toyozawa}   
\begin{eqnarray}
  \epsilon_{2} (\omega) = 
  \mathcal{N}
  \sum_{e}
  \biggl| 
  \bigl\langle 
  \Psi_{e {\bm K}={\bm 0}} 
  \bigl| 
  {\bm X} 
  \bigr| 
  \Phi_{{\rm HF}} 
  \bigr\rangle 
  \biggr|^{2}
  \delta (\Delta E_{e {\bm K}={\bm 0}} - \omega),  
  \label{eq:ISECI}  
\end{eqnarray}
 where ${\bm X}=\sum_{i} {\bm r}_{i}$ is a many-body 
 position operator of electrons, and 
 the normalization constant $\mathcal{N}$ is 
 determined via the sum rule~\cite{Ref_SUM_RULE}  
\begin{eqnarray}
 \int_{0}^{\infty} \omega \epsilon_{2} (\omega)\ d\omega = 
 \frac{\pi}{2}\omega_{p}^{2} 
  \label{eq:SUM}  
\end{eqnarray}
 with $\omega_{p}$ being the plasma frequency of the system. 
 Substituting Eq.~(\ref{eq:SECI}) into Eq.~(\ref{eq:ISECI}) and 
 noting the commutation relation 
 $\left[ \mathcal{H}_{{\rm HF}} , {\bm X} \right]  
 =   \sum_{i} \left[\hat{h}_{{\rm HF}}(i) , {\bm r}_{i} \right]
 = - \sum_{i} \partial / \partial{\bm r}_{i}$, 
 we obtain 
\begin{eqnarray}
  \epsilon_{2}(\omega) &=& 
  \mathcal{N}
  \sum_{e}
  \biggl| 
  \sum_{{\bm k}} \sum_{a}^{occ} \sum_{r}^{vir} \tilde{C}_{a r {\bm k}}^{e {\bm K}={\bm 0}} 
  \frac
  {\bigl\langle 
   \phi_{r {\bm k}} 
   \bigl| 
   \partial / \partial {\bm r} 
   \bigr| 
   \phi_{a {\bm k}} 
   \bigr\rangle }
   {\epsilon_{r {\bm k}} - \epsilon_{a {\bm k}}} 
  \biggr|^{2} \nonumber \\ 
  &\times& \delta (\Delta E_{e {\bm K}={\bm 0}} - \omega),           
  \label{eq:ISECI_2}  
\end{eqnarray}
 where $|\phi_{a {\bm k}}\rangle$ is the Bloch state 
 being the eigenstate of the Fock operator $\hat{h}_{{\rm HF}}$ and  
 the matrix element of $\partial / \partial {\bm r}$ 
 can be calculated 
 with an interpolation 
 scheme~\cite{Ref_WI} based on the MLWF's 
 from first principles (see appendix B).   
 Theoretically, 
 $\epsilon_{2}(\omega)$ 
 contains an electron-hole-interaction effect 
 due to the presence of the CI coefficients in Eq.~(\ref{eq:ISECI_2}).  
 To see the electron-hole-interaction effect on the spectrum, 
 it is convenient to compare with the spectrum obtained
 by the independent-particle approximation~\cite{Ref_IPA} (IPA) 
 where the CI coefficients are neglected in
 Eq.~(\ref{eq:ISECI_2}) and the excitations are described just 
 with optical transitions between independent
 hole and electron states,       
\begin{eqnarray}
  \epsilon_{2}^{(0)}(\omega) &=& 
  \mathcal{N}
  \sum_{{\bm k}} 
  \sum_{a}^{occ} 
  \sum_{r}^{vir} 
  \biggl| 
  \frac
  {\bigl\langle 
   \phi_{r {\bm k}} 
   \bigl| 
   \partial / \partial {\bm r} 
   \bigr| 
   \phi_{a {\bm k}} 
   \bigr\rangle }
   {\epsilon_{r {\bm k}} - \epsilon_{a {\bm k}}} 
  \biggr|^{2} \nonumber \\ 
  &\times& \delta(\epsilon_{r {\bm k}} - \epsilon_{a {\bm k}} - \omega).   
  \label{eq:IIPA}  
\end{eqnarray}

 In the spectral calculations, we have chosen a $k$-grid 
 different from the Monkhorst-Pack $k$-grid
 used in the band 
 calculations.\cite{Ref_Louie} 
 The present $k$-grid is generated as follows:  
 We first make a uniform $k$-grid in an $11 \times 11 \times 11$ 
 Monkhorst-Pack mesh and then 
 slightly shift uniformly the sampling ${\bm k}$ in the direction of 
 $-0.01 {\bm b}_{1} 
  -0.02 {\bm b}_{2} 
  +0.03 {\bm b}_{3}$ 
 with \{${\bm b}_1, {\bm b}_2, {\bm b}_3$\} being basic 
 reciprocal lattice vectors.  
 The resulting $k$-points are different from the 
 high-symmetry directions of the crystal  
 and therefore are not connected via rotational operations of the crystal
 with each other.  
 This leads to a finer sampling for the spectral calculation. 
 An unshifted grid corresponds to only 56 crystallographically 
 different points, which are too few to achieve a good spectral resolution.
 On the other hand, the shifted grid leads to a grid of 1331 
 crystallographically different $k$-points, which gives a good spectral 
 resolution. 

\begin{figure}[h]
\includegraphics[width=0.5\textwidth]{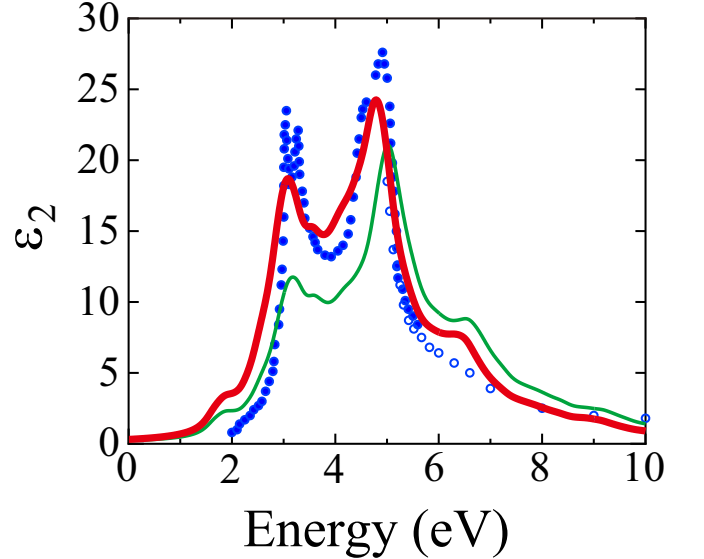}
\caption{(Color online) Calculated SECI (thick red line) and IPA (thin green line) 
 optical absorption spectrum of GaAs.
 The closed blue and open blue circles represent experimental data taken from
 Ref.~31 and Ref.~32, respectively.} 
\label{spectrum} 
\end{figure}

 We show in Fig.~\ref{spectrum} the calculated SECI (thick red line)   
 and IPA (thin green line) spectra.
 The closed blue and open blue circles denote
 experimental results.\cite{Ref_OA_EXPT_1,Ref_OA_EXPT_2}
 We see a clear contrast between the SECI and IPA spectra;  
 by considering the 
 electron-hole interaction with the SECI method,
 the spectral intensity in the low-energy region
 ($\le 5$ eV) is enhanced,         
 thus reproducing the experimental results perfectly.  
 The agreement is indeed somewhat surprising,
 when we consider the several simplified treatments employed here         
 such as the reduction of the electron-electron interaction to the extended Hubbard form.
 However, we emphasize that the nature such as the spectral enhancement 
 observed in proceeding from IPA to SECI 
 is consistent with highly-accurate full 
 {\em ab initio} results~\cite{Ref_Louie,Ref_OA_THEORY} obtained from solving 
 the Bethe-Salpeter equations. 
 In the context of the downfolding,
 these results strongly support that our model construction by the downfolding
 described in Sec.~II~C offers a reliable description
 not only of the ground-state band structure of an insulator  
 but also of the excitation spectra. 

 For the completeness,
 we perform a more critical and elaborate assessment of the 
 reliability of the downfolding; 
 we examine a sensitivity of the spectra to choices of Hamiltonian parameters.
 In the present analysis, we focus on the check of the reliability
 of the offsite interaction parameter $V$. 
 In fact, the magnitude of this parameter 
 is expected to crucially control the strength of the 
 electron-hole interaction [see Eqs.~(\ref{eq:EX}) and (\ref{eq:DR})]        
 and thus directly affect the profile of the 
 optical spectra. 
 We may calculate the excitation spectra by using choices of 
 interaction parameters different from the downfolded realistic values.
 Thus, the reliability of the downfolding can be assessed by examining
 whether the spectrum obtained from the present downfolded Hamiltonian
 gives the best agreement with the experiment among wider alternative choices of 
 the interaction parameters. 
 For this check, we introduce a scaling factor $x$ to scale 
 $V$ to $x V$. 
 With this definition, $x = 1$ corresponds to the original 
 {\em ab initio} $V$ value, while in the region $x >$ 1,
 the nearest neighbor 
 electron-electron repulsion is artificially overestimated. 
 In the practical calculation, we monitor the values of
 $x$ at 0.8, 1.0, and 1.2 and then        
 perform the SECI calculations to obtain the optical 
 spectra for each $V$ value. 
 (In the calculations, the onsite parameters 
 are fixed at the {\em ab initio}
 determined values displayed
 in TABLE~\ref{PARAM}.)
 
\begin{figure}[h]
\includegraphics[width=0.50\textwidth]{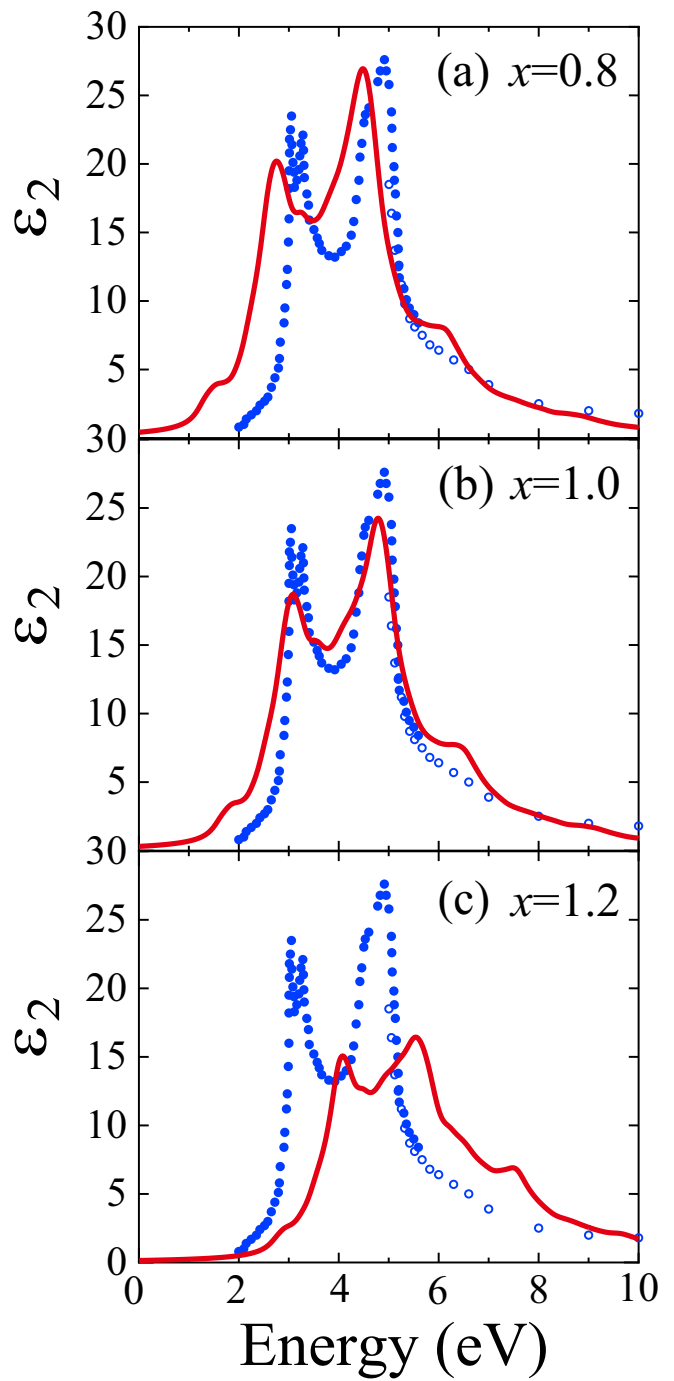}
\caption{(Color online) Dependence of SECI spectra (red lines) on scaling factor $x$;   
 (a) $x$ = 0.8, (b) $x$ = 1.0, and (c) $x$ = 1.2. 
 The closed blue and open blue circles represent experimental data taken from
 Ref.~31 and Ref.~32, respectively.} 
\label{fig:xdep} 
\end{figure}

 We show in 
 Fig.~\ref{fig:xdep}
 the resulting dependence of the SECI spectra (red lines) on 
 the scaling factor $x$. 
 The blue circles denote measured data. 
 We see a notable change in the spectra due to the parameter increase 
 [(a) $\to$ (b) $\to$ (c)];  
 increasing $x$              
 makes a blue shift 
 and an intensity decrease
 in the calculated spectra.  
 We see that the spectrum          
 at the downfolded choice (i.e., the case with $x$ = 1.0)
 exhibits the best agreement with the experiments among all the choices.
 The downfolded value offers the most realistic and accurate choices as the 
 model parameters, and thus  
 we conclude that the optical excitation
 spectrum is correctly captured by the
 downfolded Hamiltonian. 
 
 Finally, we remark a recent development for {\em ab initio}
 evaluations for the offsite $V$ parameter.
 Indeed, 
 applications of the constrained schemes to the determinations  
 of the offsite parameter is quite limited 
 in the literature compared with 
 those for the onsite parameters.  
 Recently, Aryasetiawan {\em et al}\cite{Aryasetiawan2004}
 have proposed an RPA approach for 
 calculating the interaction parameters. 
 They first calculate a real-space screened Coulomb interaction $U({\bm r}, {\bm r}')$ 
 by excluding the polarization formed in target bands 
 contained in the model Hamiltonian and then 
 evaluate the matrix element of $U$ in the localized basis 
 such as linearized muffin-tin orbitals and/or maximally-localized Wannier functions. 
 This approach is able to derive all the offsite parameters. 
 So, comparisons between the values obtained by the present constrained scheme 
 and the values based on the RPA approach would contribute to 
 making deeper understanding for the distant interaction parameters. 

\section{Conclusions}
 We have examined whether the effective low-energy Hamiltonian
 derived from the downfolding procedure is able to describe dynamics 
 and excitation spectra in a proper way.
 The calculation is performed in the three-stage scheme.
 In the first stage, we calculate the global electronic structure from
 the density functional theory supplemented by 
 the generalized gradient approximation.
 The high-energy degrees of the freedom in the global electronic band structure  
 are, in the second stage,  
 eliminated by the downfolding scheme, which leaves only the low-energy bands near the Fermi
 level. In the present example of GaAs, we retain up to 25 Ry for the calculation of the 
 global electronic bands, while the downfolded Hamiltonian keeps only eight bands near the Fermi level
 up to 15 eV ($\sim$ 1 Ry). 
 By the downfolding, kinetic and interaction energies are separately renormalized into the 
 low-energy eight bands and the effective Hamiltonian,  
 where we employ the CNDO model neglecting the offdiagonal part of the Coulomb interaction, 
 is constructed from first principles, with the help of the maximally localized Wannier functions.   
 This procedure, though several simplified treatments are employed, in principle, does not contain any
 {\it ad hoc} parameters.
 In the third stage, 
 the Hartree-Fock method for the ground state supplemented by
 the single-excitation configuration-interaction treatment
 for considering the electron-hole interactions 
 has been applied to obtain electronic excitation spectra of 
 semiconductor GaAs. 
 The spectra thus obtained have quite well 
 reproduced the experimental results; 
 the intensity and position for the excitonic peak are
 well reproduced at a quantitative level. 
 We believe that the present model construction based on the downfolding 
 offers a reliable {\em ab initio} scheme, where 
 the downfolded effective
 Hamiltonian is capable  
 of not only the ground state properties 
 but also the excitation spectra. 

 The present result opens a way of treating excitations such as the optical spectra
 by the hybrid method combining the density functional approach and the accurate
 low-energy solver for the low-energy effective models. 
 Beyond the present application to semiconductors, it would be interesting to apply this approach 
 to excitations in strongly correlated electron systems 
 such as transition metal oxides including the cuprates. 
 In the present paper, we have used the Hartree-Fock approximation
 for the ground state and the single-excitation configuration-interaction treatment 
 for the excitations.
 Optical excitation spectra of GaAs have satisfactorily 
 been treated by these approximations and the experimental 
 results have been well reproduced. 
 However, stronger electron correlation effects 
 require more sophisticated low-energy solver than 
 the Hartree Fock/single excitation configuration interaction treatment.  
 For more different and challenging issues of the electron correlation, 
 the low-energy effective Hamiltonian may indeed be treated
 by much more reliable low-energy solver for electrons in solids,
 such as quantum Monte Carlo methods for lattice Fermions,\cite{Ref_MC}             
 path-integral renormalization group method,\cite{Ref_LDA+PIRG}           
 and cluster extensions of the dynamical
 mean field theory.\cite{Ref_LDA+DMFT}   
 
 As is well known, there are many direct {\em ab initio} schemes aiming at  
 considering correlation effects; for example,
 the GW,\cite{Ref_GW,Ref_SCISSORS}
 transcorrelated,~\cite{Ref_TC} 
 and quantum Monte Carlo methods.\cite{Ref_DQMC}
 They are straightforward ways for approaching the problem   
 compared to the present approach. 
 However, the straightforward methods are faced with two serious problems:
 One is that the computational load becomes extremely heavy when all the 
 electrons or even all the valence electrons are treated equally.
 The other problem is that the so far developed straightforward methods do not
 offer a sufficiently accurate framework if the electron correlation becomes 
 strong such as in the genuine Mott insulator.  The crucial point is that 
 we need to treat dynamical as well as spatial correlations and fluctuations 
 near the Fermi level in a controllable way.
 In the present stage of the computer power, such sufficient accuracies are 
 undertaken only within simple models, which can be achieved in the low-energy
 effective model after downfolding. In fact, the high accuracy required from the 
 temporal and spatial quantum fluctuations is important only in the low-energy region
 near the Fermi level, which justifies to restrict the high-accuracy treatment only in
 the region of low-energy excitations and thus only within the downfolded Hamiltonian. 
 Within the present computer power, this downfolding procedure opens an avenue of  
 studying highly correlated electron systems as well as excitations without relying on
 {\it ad hoc} parameters.  By explicitly considering the energy hierarchy in the electronic structure,
  the first principles calculations become tractable even when 
 the electron correlation is essential.  

\begin{acknowledgments}
 We would like to thank Takashi Miyake, Atsushi Yamasaki, 
 Youhei Yamaji, Takahiro Misawa, and Taichi Kosugi for 
 helpful discussions and comments.         
 This work was supported by a Grant-in-Aid for 
 Scientific Research in Priority Areas, 
 ``Development of New Quantum Simulators 
 and Quantum Design'' (No. 17064004) of the Ministry 
 of Education, Culture, Sports, Science and Technology, Japan. 
 One of us (K.N.) acknowledges Research Fellowships of 
 the Japan Society for the Promotion of 
 Science for Young Scientists.              
 All the calculations were performed 
 on Hitachi SR11000 system 
 of the Super Computer Center at the Institute for 
 Solid State Physics, the University of Tokyo.          
\end{acknowledgments}

\appendix
\section{Derivation of the fitting function of Eq.~(\ref{eq:fitting})}
 Here, we describe the details of the fitting function used in 
 the onsite-parameter determination.
 In the atomic limit, 
 the onsite Hamiltonian for the decoupled site $I$ in the home cell 
 is written as 
\begin{eqnarray}
  \mathcal{H}_{C}^{I {\bm 0}}
 &=& 
  U_{I} \sum_{\mu} 
  N_{\mu I {\bm 0}}^{\uparrow} N_{\mu I {\bm 0}}^{\downarrow}  
+ U'_{I}\sum_{\mu < \nu}
  N_{\mu I {\bm 0}} N_{\nu I {\bm 0}} 
+ \epsilon_{I} \sum_{\mu} 
  N_{\mu I {\bm 0}}  \nonumber \\ 
&=& 
  \frac{U_{I}}{2} \sum_{\mu} 
  N_{\mu I {\bm 0}} \left( N_{\mu I {\bm 0}}-1 \right)  
+ \frac{U'_{I}}{2} \sum_{\mu \ne \nu}
  N_{\mu I {\bm 0}} N_{\nu I {\bm 0}} \nonumber \\ 
&+& \epsilon_{I} \sum_{\mu} 
  N_{\mu I {\bm 0}},    
  \label{eq:H_onsite}  
\end{eqnarray}
 where $N_{\mu I {\bm 0}}^{\sigma}$ and  
 $N_{\mu I {\bm 0}}=\sum_{\sigma} N_{\mu I {\bm 0}}^{\sigma}$
 are number operators.
 $U_{I}$ and $U'_{I}$ are  
 the onsite intraorbital 
 and interorbital Coulomb repulsions, respectively. 
 $\epsilon_{I}$ is a chemical potential.
 The onsite energy $E_{C}^{I {\bm 0}}$ derived from Eq.~(\ref{eq:H_onsite}) 
 is expressed with the atomic-limit wave function $\left| \Phi_{AL} \right> $ as 
\begin{eqnarray}
 E_{C}^{I {\bm 0}} &=& 
  \langle \Phi_{AL} \left| 
  \mathcal{H}_{C}^{I {\bm 0}}
  \right| \Phi_{AL} \rangle \nonumber \\ 
&=& 
  \frac{U_{I}}{2} \sum_{\mu} 
  q_{\mu I {\bm 0}} \left( q_{\mu I {\bm 0}}-1 \right)  
+ \frac{U'_{I}}{2} \sum_{\mu \ne \nu}
  q_{\mu I {\bm 0}} q_{\nu I {\bm 0}} \nonumber \\ 
&+& \epsilon_{I} \sum_{\mu} 
  q_{\mu I {\bm 0}}, 
  \label{eq:E_C}  
\end{eqnarray}
 where we used  
 $ N_{\mu I {\bm 0}} \left| \Phi_{AL} \right>=   
   q_{\mu I {\bm 0}} \left| \Phi_{AL} \right>$.   
 We introduce $\delta q_{\mu I {\bm 0}} = q_{\mu I {\bm 0}}
 - \bar{q}_{I{\bm 0}}$ with 
 $\bar{q}_{I{\bm 0}}$ defined as the orbital occupancy
 at the equilibrium state.
 At the equilibrium state,
 the term linear in $\delta q_{\mu I {\bm 0}}$   
 should vanish, which results in cancellation of  
 the chemical potential term with 
 the linear term from the Coulomb contribution.
 Then the quadratic energy difference due to the charge fluctuations is derived as 
\begin{eqnarray}
 \Delta E_{C}^{I {\bm 0}} = \frac{U_{I}}{2} \sum_{\mu} \left( \delta q_{\mu I {\bm 0}} \right)^{2}
  + \frac{U'_{I}}{2} \sum_{\mu \ne \nu} \delta q_{\mu I {\bm 0}} \delta q_{\nu I {\bm 0}}.  
 \label{eq:U_sym} 
\end{eqnarray}
 By noting $\sum_{\mu} \left( \delta q_{\mu I {\bm 0}} \right)^{2} =
            \left( \sum_{\mu} \delta q_{\mu I {\bm 0}} \right)^{2} 
          - \sum_{\mu \ne \nu} \delta q_{\mu I {\bm 0}} \delta q_{\nu I {\bm 0}}$ and 
 defining the site-charge fluctuation
 $\delta Q_{I {\bm 0}} = \left( \sum_{\mu} \delta q_{\mu I {\bm 0}} \right)$, we obtain the form 
\begin{eqnarray}
 \Delta E_{C}^{I {\bm 0}}=\frac{U_{I}}{2} \left( \delta Q_{I {\bm 0}} \right)^{2}
  +\frac{1}{2} \left( U'_{I} - U_{I} \right)
 \sum_{\mu \ne \nu} \delta q_{\mu I {\bm 0}} \delta q_{\nu I {\bm 0}}.  
 \label{eq:E_SITE} 
\end{eqnarray}
 One may specialize the charge fluctuation of one orbital because of 
 the crystallographycal symmetry in the system.  
 Thus, the orbital index in the orbital-charge fluctuation is dropped and  
 the cross term in Eq.~(\ref{eq:E_SITE}) is rewritten in terms of $\delta Q_{I {\bm 0}}$
 and $\delta q_{I {\bm 0}}$, 
\begin{eqnarray}
  \sum_{\mu \ne \nu} \delta q_{\mu I {\bm 0}} \delta q_{\nu I {\bm 0}}
 = 4 \delta q_{I {\bm 0}} \left( \delta Q_{I {\bm 0}} - \delta q_{I {\bm 0}} \right).  
 \label{eq:CROSS} 
\end{eqnarray}
 Inserting the above expression into Eq.~(\ref{eq:E_SITE}) leads to the following 
 expression for the onsite interaction energy:
\begin{eqnarray}
 \Delta E_{C}^{I {\bm 0}} &=& \frac{U_{I}}{2} \left( \delta Q_{I {\bm 0}} \right)^{2}
  + 2 \left( U'_{I} - U_{I} \right) \delta Q_{I {\bm 0}} \delta q_{I {\bm 0}} \nonumber \\ 
  &+& \left[ 2 \left( U_{I} - U'_{I} \right) \right] \left( \delta q_{I {\bm 0}} \right)^{2}.
\end{eqnarray}

\section{Evaluation of the matrix element of $\partial / \partial {\bm r}$}
 Here we describe details for the calculation of 
 the matrix element $\bigl\langle 
   \phi_{r {\bm k}} 
   \bigl| 
   \partial / \partial {\bm r} 
   \bigr| 
   \phi_{a {\bm k}}  
   \bigr\rangle$ 
 in Eq.~(\ref{eq:ISECI_2}). 
 We first rewrite it in terms of the 
 maximally localized Wannier function as         
\begin{eqnarray}
   \biggl\langle 
   \phi_{r {\bm k}} 
   \biggl| 
   \frac{\partial}{\partial {\bm r}}  
   \biggr| 
   \phi_{a {\bm k}} 
   \biggr\rangle &=&  
   \frac{1}{N} 
   \sum_{\mu \nu} \sum_{i j} \sum_{{\bm R} {\bm R}'} 
   C_{\mu i}^{r *} ({\bm k}) 
   C_{\nu j}^{a  } ({\bm k}) 
   e^{i {\bm k} \cdot ({\bm R}'-{\bm R})} \nonumber \\ 
   &\times& \biggl\langle 
   w_{\mu i {\bm R}} 
   \biggl| 
   \frac{\partial}{\partial {\bm r}}  
   \biggr| 
   w_{\nu j {\bm R}'} 
   \biggr\rangle   
  \label{eq:wdrw}  
\end{eqnarray}
 with 
  $\bigl| 
  \phi_{\alpha \bk}
  \bigr\rangle =
  \left( 1/ \sqrt{N} \right)  
  \sum_{\mu i} C_{\mu i}^{\alpha}(\bk) 
  \sum_{{\bm R}} e^{i {\bm k}\cdot{\bm R}}
  \bigl| 
  w_{\mu i {\bm R}}
  \bigr\rangle$. 
 In our calculation, the Wannier function 
  $ w_{\mu i}({\bm r}-{\bm R}) =  
  \bigl\langle {\bm r} 
  \bigr| 
   w_{\mu i {\bm R}} 
  \bigr\rangle$ is stored as numerical data on the real-space grid, 
\begin{eqnarray}
  {\bm r} = \frac{m_1}{M_1}{\bm L}_1
          + \frac{m_2}{M_2}{\bm L}_2 
          + \frac{m_3}{M_3}{\bm L}_3,  
  \label{eq:vec_r}  
\end{eqnarray}
 where ${\bm L}_{i}$($=N_{i} {\bm a}_{i}$) is a cell vector 
 defining a superlattice containing $N(=N_1 N_2 N_3)$ 
 primitive cells,           
 $m_i$ runs on the integer values: $0, 1, \cdots, M_{i}-1$ with $M_i$ being the total 
 number of the grids in the $i$th direction       
 (in the present case, $M_{1}=M_{2}=M_{3}=240$).   
 Since the Wannier function satisfies the following periodic 
 boundary condition, 
\begin{eqnarray}
  w_{\mu i}({\bm r}+{\bm L}_{i}) =  
  w_{\mu i}({\bm r}),\ \ \ \ \ \ i=1,2,3,     
  \label{eq:pbc}  
\end{eqnarray}
 we can express $w_{\mu i}({\bm r})$ in terms of the 
 Fourier transformation as  
\begin{eqnarray}
  w_{\mu i}({\bm r}) =  
  \sum_{{\bm G}_{L}} 
  w_{\mu i}({\bm G}_{L}) e^{i {\bm G}_{L} \cdot {\bm r}}, 
  \label{eq:FT}  
\end{eqnarray}
 where ${\bm G}_{L} 
 = g_{1} {\bm L}_{1}^{*}  
 + g_{2} {\bm L}_{2}^{*}  
 + g_{3} {\bm L}_{3}^{*}$ with  
 $ {\bm L}_{i}^{*}=(2\pi/V){\bm L}_{j} \times {\bm L}_{k}$, 
 $V=({\bm L}_{1} \cdot {\bm L}_{2} \times {\bm L}_{3})$ is 
 the volume of the superlattice, and 
 $g_i$ takes value from $1$ to $M_{i}-1$.
 We note that ${\bm G}_{L}$ is different from ${\bm G}$ used in the 
 {\em ab initio} band calculations; 
 the former is expressed in terms of the reciprocal-lattice vectors for 
 the superlattice 
 \{${\bm L}_{1}^{*}, 
    {\bm L}_{2}^{*}, 
    {\bm L}_{2}^{*}$\}, 
 while the latter is written with the basic reciprocal lattice vectors 
 \{${\bm b}_1, {\bm b}_2, {\bm b}_3$\}.
 Substituting Eq.~(\ref{eq:FT}) into Eq.~(\ref{eq:wdrw}) and 
 using the translational symmetry for the matrix element leads to 
\begin{eqnarray}
   \biggl\langle 
   \phi_{r {\bm k}} 
   \biggl| 
   \frac{\partial}{\partial {\bm r}}  
   \biggr| 
   \phi_{a {\bm k}} 
   \biggr\rangle &=& 
   \sum_{\mu \nu} \sum_{i j} 
   C_{\mu i}^{r *} ({\bm k}) 
   C_{\nu j}^{a  } ({\bm k}) \nonumber \\ 
   &\times& \sum_{{\bm R}} 
   g_{\mu i \nu j} ({\bm R}) 
   e^{i {\bm k} \cdot {\bm R}} 
  \label{eq:wdrw_2}  
\end{eqnarray}
 with     
\begin{eqnarray}
   g_{\mu i \nu j} ({\bm R}) = 
   i V \sum_{{\bm G}_{L}}
   w_{\mu i}^{*} ({\bm G}_{L}) 
   w_{\nu j} ({\bm G}_{L}) 
   {\bm G}_{L} 
   e^{-i {\bm G}_{L} \cdot {\bm R}}.                 
  \label{eq:gij}  
\end{eqnarray}
 The actual calculation proceeds as follows: 
 We first transform the real-space Wannier function $w_{\mu i}({\bm r})$
 into the reciprocal-space one $w_{\mu i}({\bm G}_{L})$ in Eq.~(\ref{eq:FT}) with 
 the algorithm of the fast-Fourier-transformation with radix-2, 3, and 5. 
 Then, we calculate the $\partial / \partial {\bm r}$ matrix in the 
 Wannier basis [${\bm g}({\bm R})$ in Eq.~(\ref{eq:gij})] to obtain 
 the desired quantity [Eq.~(\ref{eq:wdrw_2})]. 
 We note that this numerical procedure is a so-called 
 Wannier interpolation scheme;\cite{Ref_WI}  
 we construct the Wannier functions with the {\em ab initio} Bloch 
 functions in 
 the Monkhorst-Pack 
 $k$-grid 
 and then 
 interpolate the matrix elements of $\partial / \partial {\bm r}$ 
 at the slightly shifted 
 $k$-grid used in the 
 spectral calculation.


\begin{references}
\bibitem{Ref_supercell}
 Our constrained calculations were performed for 
 a 3$\times$3$\times$3 fcc supercell containing 
 54 atoms, with 7$\times$7$\times$7 $k$-points sampling. 
 It was found that the sizes of the supercell and sampling $k$-points 
 are sufficient for obtaining a converged interaction parameters.
\end{references}
\end{document}